\documentclass[aps,preprint,superscriptaddress,]{revtex4}
\usepackage{amsfonts}
\usepackage{amsmath}
\usepackage{amssymb}
\usepackage{graphicx}

\begin{document}

\title{Information conservation is fundamental: recovering the lost
information in Hawking radiation}
\author{Baocheng Zhang}
\email{zhangbc@wipm.ac.cn}
\affiliation{State Key Laboratory of Magnetic Resonances and Atomic and Molecular
Physics, Wuhan Institute of Physics and Mathematics, Chinese Academy of
Sciences, Wuhan 430071, People's Republic of China}
\author{Qing-yu Cai}
\email{qycai@wipm.ac.cn}
\affiliation{State Key Laboratory of Magnetic Resonances and Atomic and Molecular
Physics, Wuhan Institute of Physics and Mathematics, Chinese Academy of
Sciences, Wuhan 430071, People's Republic of China}
\author{Ming-sheng Zhan}
\email{mszhan@wipm.ac.cn}
\affiliation{State Key Laboratory of Magnetic Resonances and Atomic and Molecular
Physics, Wuhan Institute of Physics and Mathematics, Chinese Academy of
Sciences, Wuhan 430071, People's Republic of China}
\author{Li You}
\email{lyou@mail.tsinghua.edu.cn}
\affiliation{State Key Laboratory of Low Dimensional Quantum Physics, Department of
Physics, Tsinghua University, Beijing 100084, China}

\begin{abstract}
In both classical and quantum world, information cannot appear or disappear.
This fundamental principle, however, is questioned for a black hole, by the
acclaimed \textquotedblleft information loss paradox". Based on the
conservation laws of energy, charge, and angular momentum, we recently show
the total information encoded in the correlations among Hawking radiations
equals exactly to the same amount previously considered lost, assuming the
non-thermal spectrum of Parikh and Wilczek. Thus the information loss paradox
can be falsified through experiments by detecting correlations, for instance,
through measuring the covariances of Hawking radiations from black holes,
such as the manmade ones speculated to appear in LHC experiments.
The affirmation of information conservation in Hawking radiation will shine
new light on the unification of gravity with quantum mechanics.

\end{abstract}

\pacs{03.67.-a, 04.70.Dy, 04.60.-m}
\maketitle

Information is physical \cite{lr91}, it cannot simply disappear in any
physical process. This basic principle of information science constitutes
one of the most important elements for the very foundation to our daily life
and to our understanding of the universe. Unfortunately, it is in stark
confrontation with the information loss paradox, which claims information
can be lost in a black hole, allowing many distinct physical states to
evolve into the same final state.

So what is information? According to the standard textbook definition,
information refers to an inherent property concerning the amount of
uncertainty for a physical system. Shannon provided a mathematical formula
to quantify and measure information given the state of a system \cite%
{shannon}. Classically, a physical state is specified by a distribution
function in the multi-dimensional phase space for all its degrees of
freedom. This distribution evolves according to Liouville's theorem, which
conserves the phase space volume and gives rise to the conservation of
entropy or information under Hamiltonian dynamics. Classical information is
stored, processed, and communicated according to classical physics. In
parallel, quantum information follows the laws of quantum mechanics. For
either a pure state specified by a wave function or a mixed state specified
by a density matrix, its quantum information content is measured by von
Neumann entropy instead \cite{nc00,sl05}. In quantum physics,
information is physical takes on a more demanding role. For instance, we
know that quantum information can neither be cloned \cite{noclone1}, nor
deleted \cite{Pati} or split \cite{Zhou}.

Shannon's entropy, is analogous to thermodynamic entropy, which measures the
degree of uncertainty in a physical system. Perhaps the example of a
classical bit of information in a modern computer device studied by Landauer
\cite{rl61} best illustrates this point. Consistent with the second law of
thermodynamics, the entropy for the environment increases after the erasure
of the bit of information because useful work is consumed in the process.
Entropy (or information) for the whole system of the bit encoding device plus
its environment, nevertheless, remains conserved \cite{cai04}.

The second example concerns the famous Maxwell's demon \cite{max18,ls29},
who paradoxically prepares a more ordered state (with less entropy) by
performing microscopic operations to each individual atoms in a gas
ensemble. Its resolution comes from the realization of the demon's ever
increasing amount of information gained in order to properly act on
each passing atom. Given the demon's finite capacity to store
information, he therefore must update stored information with newly acquired
information, consequently increasing entropy for the environment \cite{nc00}.
Yet, the entropy or information remains conserved for the total system of the
demon, the atoms, and the environment.

The third example concerns quantum teleportation \cite{bbc93}, the paradigm
of quantum information science whereby an unknown quantum state is
disassembled at the sender and regenerated at the receiver. It was once
thought that faster than the speed of light communication may become
possible when teleportation is employed \cite{nc00}. However,
according to the special theory of relativity, no physical system can travel
faster than light, including the information encoded in an unknown physical
state as information itself is physical.
During quantum teleportation, the reconstruction of the unknown initial
state at the receiving end requires a classical channel
communicating to the sender, which is ultimately limited by the speed of light.

The above examples illustrate that information is conserved for a closed
system, be it classical or quantum.
The discovery of Hawking radiation from a black hole \cite{swh74,swh75},
however, brings up a serious challenge to the principle of information
conservation. While controversial, the conjecture that information could
disappear, or is not conserved in a black hole, is nevertheless supported by
its share of believers. This so-called paradox of black hole
information loss contradicts with fundamental laws of physics that
support conservation of entropy or information.

Historically, a black hole was first viewed as only capable of absorbing but
not emitting particles according to classical physics. This was not
considered to contradict information conservation principle. As particles
disappear into a black hole, the information encoded remains contained
inside the black hole, although inaccessible to outside observers.
Thus this information is simply considered hidden inside a black
hole, not a real loss to the outside. However, the contradiction arises
after the discovery of Hawking radiation \cite{swh74,swh75}
according to the theory of relativity, quantum mechanics, and thermodynamics.

Hawking asserted that the emitted radiation from a black hole is thermal and
its detailed form is independent of the structure of matter that collapsed
to form a black hole \cite{swh76}. The radiation state is thus considered
a completely mixed one which is incapable of carrying information about how
a black hole is formed. The fact that radiation outside a black hole is in
a mixed or a thermal state is in itself not in contradiction with quantum
mechanics. When a system is composed of several parts, the reduced state for
any part or a subsystem may be mixed, and typically the added information
from all parts decreases. We understand the mechanism for this information loss is quantum
entanglement. When two subsystems are entangled, their respective entropies
increase, equivalent to a decease of information contained in
the parts. The apparent loss to the total system information
calculated from summing up the information from its two subsystems is due to
information hidden inside correlations between the subsystems. In the
case of a radiating black hole, it is the correlation between the outgoing
radiation states and the internal inaccessible states of a black hole.
According to our understanding, a rudimentary form
of the black hole information loss paradox refers to the following:
after a black hole is exhausted into Hawking radiations, its
complete thermal state of Hawking radiations contains no information at all.

In a recent article, the possibility that information about infallen matter
could hide in the correlations between the Hawking radiation and the
internal states of a black hole is ruled out \cite{bp07}. This made the
information loss paradox even more severe, \textit{i.e.}, the paradox arises
immediately after a black hole starts to emit thermal Hawking radiations.

We have followed up, recently visited and revisited this same problem \cite%
{zcyz09,zcyz}. In our opinion, we have provided a satisfactory resolution
based on the fundamental principles of physics and statistical theory. At
the heart of our proposal is the discovery by us of the existence of
correlations among Hawking radiations. Our calculations show that the amount
of information encoded in this correlation exact equals to the same amount of
information claimed lost by the information loss paradox \cite{zcyz09,zcyz}.
In this essay, we present our current understanding of the information
conservation principle in the context of Hawking radiation as tunneling
\cite{pw00}. Additionally, we provide further support to our proposal
by computing the ratio for entropy production and show
in detail how the lost information is balanced by the correlation hidden in
Hawking radiation.

The original treatment of Hawking radiation \cite{swh74,swh75} gives
a thermal spectrum, which is inconsistent with energy conservation, due to the
approximation of a fixed background geometry.
In contrast, the approach of
Hawking radiation as quantum tunneling, first treated by Parikh and
Wilczek \cite{pw00}, strictly enforces energy conservation for the $s$-wave
outgoing tunneled particles. Particles are supplied by the consideration of the
geometrical limit due to the infinite blue shift of the outgoing
wave-packets near horizon. The tunneling barrier is created by the
outgoing particle itself, which is ensured by energy conservation. Making
use of the Painlev\'{e} coordinate system that is regular at horizon, a
non-thermal spectrum for Hawking radiation, or the energy dependent
tunneling probability
\begin{eqnarray}
\Gamma (M;E) &\sim &\exp \left[ -8\pi E\left( M-{E}/{2}\right) \right],
\label{1tp}
\end{eqnarray}%
is found for a Schwarzschild black hole of a mass $M$ \cite{pw00}. We will
adopt the convenient units of $k=\hbar =c=G=1$. $M$ is omitted when no
ambiguity arises. This result is consistent with the change of the
Bekenstein-Hawking entropy $S_{\mathrm{BH}}$ for a black hole as shown in
Ref. \cite{pw00}, or $\Gamma (M;E)=\exp(\Delta S_{\mathrm{BH}})$. It is
clearly distinct from a thermal distribution of $\exp \left( -8\pi EM\right)
$. As a result, Hawking radiations must be correlated, and their correlations can
carry away information encoded within \cite{zcyz09,zcyz}.

How to probe the existence of this correlation? Given two statistical
events, like two emissions of Hawking radiation, with their joint
probability denoted by $p(A, B)$, the probabilities for the two respective
emissions are given by $p(A)=\int p(A,B)dB$ and $p(B)=\int p(A,B)dA$. We can
proceed with a simple check to see whether $p(A,B)=p(A)\cdot p(B)$ holds
true or not. If the equality sign holds true, then there exists no
correlation in between. The two events are statistically independent. This
is indeed the case when the emission spectrum is taken to be thermal.
Otherwise for a non-thermal spectrum such as the one given by Eq. (\ref{1tp}),
the equality sign does not hold.
The two emissions are dependent or correlated.
Alternatively, we can check if the conditional probability $%
p(B|A)=p(A,B)/p(A)$ for event $B$ to occur given that event $A$ has occurred
is equal to the probability $p(B)$ or not \cite{gs92}.

We have previously shown \cite{zcyz09,zcyz}, the joint probability $\Gamma
(E_{1},E_{2})$ for two emissions of Hawking radiation, one at an energy $%
E_{1}$ and another at an energy $E_{2}$, is given by
\begin{eqnarray}
\Gamma (E_{1},E_{2})&=&\exp \left[ -8\pi (E_{1}+E_{2})\left( M-({E_{1}+E_{2}}%
)/2\right) \right], \hskip 18pt
\end{eqnarray}
\textit{i.e.}, the probability of two emissions at energies $E_1$ and $E_2$
is precisely the same as the emission probability for a single Hawking
radiation at an energy $E_1+E_2$. This can be stated explicitly as
\begin{eqnarray}
\Gamma (E_{1},E_{2})&=&\Gamma (E_{1}+E_{2}),  \label{ec}
\end{eqnarray}
which holds true as \textit{energy conservation} is enforced within the
treatment of Hawking radiation as tunneling. The existence of correlation is thus
affirmed because we find
\begin{eqnarray}
\Gamma (E_{1},E_{2})\neq \Gamma (E_{1})\cdot \Gamma (E_{2}).
\end{eqnarray}%
Alternatively, the existence of correlation is revealed as
$\Gamma (E_{2}|E_{1})\neq \Gamma (E_{2})$, where
the conditional probability is found to be
\begin{eqnarray}
\Gamma (E_{2}|E_{1})&=&{\Gamma (E_{1},E_{2})}/{\Gamma \left( E_{1}\right) }
\notag \\
&=&\exp \left[ -8\pi E_{2}\left( M-E_{1}-{E_{2}}/{2}\right) \right],
\label{cp}
\end{eqnarray}%
for an emission with energy $E_{2}$ given the occurrence of an emission with
energy $E_{1}$.

How to measure correlation in terms of the amount of information it can
encode? A closely related topic in quantum information theory provides
useful clue in terms of mutual information between two parties.
It constitutes a legitimate correlation measure for any bipartite system.
Between two Hawking radiations with energies $E_{1}$ and $E_{2}$,
mutual information is defined according to \cite{nc00}
\begin{eqnarray}
S(E_{2}:E_{1}) &\equiv & S(E_{2})+S(E_{1})-S(E_{1},E_{2})  \notag \\
&=&S(E_{2})-S(E_{2}|E_{1}),  \label{mcd}
\end{eqnarray}%
where $S(E_{1},E_{2})$ is the entropy for the system and $S(E_{2}|E_{1})$ is
the conditional entropy. A few simple substitutions then yields $%
S(E_{2}:E_{1})=-\ln \Gamma (E_{2})+\ln \Gamma (E_{2}|E_{1}) = 8\pi E_{1}E_{2}
$ between the two Hawking radiations, confirming what we discovered earlier:
\textit{there exist hidden messengers in Hawking radiation} \cite{zcyz09}.

For a queue of emissions $E_i$ ($i=1,2,\cdots,n$), the total correlation can
be calculated along any one of the independent partitions \cite{zcyz09,zcyz}.
We choose a partition according to the queue of the subscripts, expressing
the total correlation as the sum of the correlations between emissions $E_1$
and $E_2$, $E_1\oplus E_2$ and $E_3$, $E_1\oplus E_2\oplus E_3$ and $E_4$, $%
\cdots$, and $E_1\oplus E_2\cdots \oplus E_{n-1}$ and $E_n$, where
$E_A\oplus E_B$ denotes the combined system of $E_A$ and $E_B$.
Such a partition allows us to easily compute the total correlation,
which is different from simply summing up correlations
between pairs of emissions $E_i$ and $E_{j\neq i}$.
When the total correlation among all emissions is obtained, the amount of information
it can encode is found again to be exactly equal to the amount previously
considered lost by the information loss paradox \cite{zcyz09,zcyz}. Thus,
while its properties are fascinating, a black hole is nothing fundamentally
special when it comes to the principle of information conservation. It remains a
physical system governed by the laws of physics we are accustomed to,
including the conservation of information.

If Hawking radiation were thermal, its entropy is previously found to be
approximately equal to ${4}/{3}$ times the reduced amount of the entropy for a
Schwarzschild black hole \cite{zw82,page83}. Expressed in terms of entropy
production ratio, this becomes $R=dS^{^{\prime }}/dS_{\mathrm{BH}}\simeq {4}/%
{3}$, where $dS^{^{\prime }}$ denotes the change of entropy for the thermal
radiation. The increase of entropy implies the loss of information, thus the
information loss paradox.

When the spectrum is non-thermal, an extra term in the entropy $S\left(
E_{f}|E_{i}\right) =-\ln \Gamma (E_{f}|E_{i})$ arises from
the correlation between the two emissions \cite{zcyz09}, where $\Gamma
(E_{f}|E_{i})=\exp \left[ -8\pi E_{f}\left( M-E_{i}-{E_{f}}/{2}\right) %
\right] $ is the conditional probability for an emission at energy $E_{f}$
given an emission of energy $E_{i}$, or conditional on the total energy of
all previous emissions being $E_i$ according to Eq. (\ref{ec}).

We now prove that the microscopic process of a Hawking radiation with an
energy $dE$ is unitary. The entropy carried by an emission $dE$ is
\begin{eqnarray}
dS=-\ln \Gamma (dE)=8\pi dE\left( M-{dE}/{2}\right),
\end{eqnarray}%
which can be compared with the increase of the Bekenstein-Hawking entropy of
a black hole
\begin{eqnarray}
dS_{\mathrm{BH}}=4\pi \lbrack (M-dE)^{2}-M^{2}]=-8\pi dE\left( M-{dE}/{2}%
\right). \hskip 12pt
\end{eqnarray}%
This gives an entropy production ratio of $R=\left|dS/dS_{\mathrm{BH}%
}\right|=1$ for the non-thermal spectrum of Eq. (\ref{1tp}). Thus
the process of Hawking radiation is thermodynamically reversible, its \textit{%
entropy or information is conserved}. The decrease of the entropy for a
black hole is exactly balanced by the increase of the entropy in its emitted
radiations.

According to Bekenstein \cite{jdb80}, there exists a one-to-one
correspondence between the internal state of a black hole and its
precollapsed configuration when its entropy is interpreted in terms of the
Boltzmann's formula. The proof of the conservation of total entropy for a
black hole and its Hawking radiation, together with the Bekenstein
conjecture, thus implies a one-to-one correspondence between the
precollapsed configuration and the state of Hawking radiation \cite{zczy11}.
Therefore the complete process of Hawking radiation is unitary.

What happens when a black hole is exhausted due to emission of Hawking radiation?
For the non-thermal spectrum of Eq. (\ref{1tp}), the entropy of Hawking radiation
can be computed through counting of the microstates denoted by $\left(
E_{1},E_{2},\cdots ,E_{n}\right)$, constrained by \textit{energy conservation} $%
\sum_{i}E_{i}=M$. Within such a description, a fixed queue of $E_{i}$
specifies a microstate. The probability for the microstate $\left(
E_{1},E_{2},\cdots ,E_{n}\right) $ to occur is given by
\begin{eqnarray}
P_{( E_{1},E_{2},\cdots ,E_{n})} &=& \Gamma (M;E_{1})\cdot \Gamma
(M-E_{1};E_{2})\cdot  \notag \\
&& \cdots \Gamma\left(M-\sum_{j=1}^{n-1}E_{j};E_{n}\right).
\end{eqnarray}%
Given the non-thermal spectrum, each of the factors on the right hand side
can be easily computed. For example, we find for the last emission $%
\Gamma(M-\sum_{j=1}^{n-1}E_{j};E_{n}) =\exp (-4\pi E_{n}^{2})$.
Collecting all factors together we obtain $P_{(
E_{1},E_{2},\cdots ,E_{n})} =\exp (-4\pi M^{2})=\exp (-S_{\mathrm{BH}})$.
According to the fundamental postulate of statistical mechanics all
microstates of an isolated system are equally likely, thus the number of
microstates $\Omega ={1}/{P_{( E_{1},E_{2},\cdots ,E_{n})} }=\exp (S_{%
\mathrm{BH}})$, and the total entropy of Hawking radiation becomes $S=\ln
\Omega =S_{\mathrm{BH}}$ using Boltzmann's definition of entropy. The total
entropy carried away by all emissions is thus shown to be precisely equal to
the entropy in the original black hole. Therefore, due to \textit{entropy
conservation}, Hawking radiation must be \textit{unitary}.

We have recently extended our resolution for the black hole information loss
paradox \cite{zcyz09} to an extensive list of other types of black holes,
including charged black holes, Kerr black holes, and Kerr-Neumann black
holes \cite{zcyz}. In their respective cases, our resolution is shown to
remain valid provided the spectra for Hawking radiation as tunneling are
non-thermal as required by the conservation laws of physics: \textit{charge
conservation, angular momentum conservation, and energy conservation}.
Even when effects of quantum gravity \cite{zcyz} and non-commutative
black holes \cite{zcy11} are involved, our discovery of correlation among
Hawking radiations remains effective in providing a consistent resolution to
the black hole information loss paradox.

After conclusively show that the non-thermal spectrum of Parikh and
Wilczek allows for the Hawking radiation emissions to carry away all
information of a black hole, a natural question arises as to
whether Hawking
radiation is indeed non-thermal or not? Although the derivation of the
non-thermal spectrum is based on solid physics, it remains to
be confirmed experimentally or in observations. A recent analysis by us show
that the non-thermal spectrum can be distinguished from the thermal spectrum by
counting the energy covariances of Hawking radiations \cite{zcy13}. With the
relatively low energy scale for quantum gravity and the large dimensions,
the production of micro black holes and the corresponding observation of Hawking
radiations has already been studied \cite{bf99,cms11,mr08,gt02,dl01,ehm00}.
If Hawking radiations from a micro black hole were observed in an LHC
experiment, our results show that it can definitely determine whether the
emission spectrum is indeed non-thermal \cite{zcy13} or not. Thus it
provides an avenue towards experimentally testing the long-standing
\textquotedblleft information loss paradox\textquotedblright .

The series of studies by us reveal that information is
conserved in an isolated system. For a black hole,
the total system includes its radiations.
The principle of information conservation remains true when
correlations among Hawking radiations are properly taken into account.
Hawking radiation is unitary,
which implies the dynamics of black hole obey the laws of quantum
mechanics. Since black hole results from Einstein's field equation,
the unitarity for a black hole definitely indicates the possibility of a unified
gravity and quantum mechanics.

{\it Acknowledgement.}---Financial support from NSFC under Grant Nos.
11074283, 11104324, and 11004116,
and NBRPC under Grant Nos. 2010CB832805, 2012CB922100,
and 2013CB922000 is gratefully
acknowledged.

\end{document}